\documentclass[reprint,twocolumn,amsmath,amssymb,superscriptaddress]{revtex4-2}
\usepackage{graphicx, color, graphpap}
\usepackage[caption=false,subrefformat=parens,labelformat=parens]{subfig}
\usepackage{amsmath}
\usepackage{overpic}
\usepackage{physics}
\usepackage{enumitem}
\usepackage{comment}
\usepackage{amssymb}
\usepackage{amsthm}
\usepackage{pstricks} 
\usepackage{float}
\usepackage{multirow}
\usepackage{hyperref}
\usepackage[T1]{fontenc}
\usepackage{framed}
\usepackage{bm}
\usepackage{bbm}
\usepackage{mathtools}
\usepackage{comment}
\usepackage{stackengine}

\definecolor{shadecolor}{gray}{0.9}

\renewcommand{\Tr}[1]{\mathrm{Tr}\left[#1\right]}

\usepackage{placeins}

\begin{document}
	
\title{Harnessing high-dimensional temporal entanglement using limited interferometric setups}

\author{Alexandra Bergmayr-Mann}
\altaffiliation{These authors contributed equally, the order was chosen randomly}
\thanks{alexandra.bergmayr@gmx.at}
\affiliation{Vienna Center for Quantum Science and Technology (VCQ), Atominstitut, Technische Universität Wien, Stadionallee 2, 1020 Vienna, Austria}

\author{Florian Kanitschar}
\altaffiliation{These authors contributed equally, the order was chosen randomly}
\thanks{florian.kanitschar@outlook.com}
\affiliation{Vienna Center for Quantum Science and Technology (VCQ), Atominstitut, Technische Universität Wien, Stadionallee 2, 1020 Vienna, Austria}
\affiliation{AIT Austrian Institute of Technology, Center for Digital Safety \& Security, Giefinggasse 4, 1210 Vienna, Austria}

\author{Matej Pivoluska}
\email{pivoluskamatej@gmail.com}
\affiliation{Vienna Center for Quantum Science and Technology (VCQ), Atominstitut, Technische Universität Wien, Stadionallee 2, 1020 Vienna, Austria}
\affiliation{Institute of Computer Science, Masaryk University, 602 00 Brno, Czech Republic}
 \affiliation{Institute of Physics, Slovak Academy of Sciences, 845 11 Bratislava, Slovakia}

\author{Marcus Huber}
\email{marcus.huber@tuwien.ac.at}
\affiliation{Vienna Center for Quantum Science and Technology (VCQ), Atominstitut, Technische Universität Wien, Stadionallee 2, 1020 Vienna, Austria}
 \affiliation{Institute for Quantum Optics and Quantum Information (IQOQI),
Austrian Academy of Sciences, Boltzmanngasse 3, 1090 Vienna, Austria}

\date{\today}

\begin{abstract}
High-dimensional entanglement has been shown to provide significant advantages in quantum communication. One of its most promising implementations is available in the time-domain routinely produced in spontaneous parametric down-conversion (SPDC). While advantageous in the sense that only a single detector channel is needed locally, it is notoriously hard to analyze, especially in an assumption-free manner that is required for quantum key distribution applications. We develop the first complete analysis of high-dimensional entanglement in the polarization-time-domain and show how to efficiently certify relevant density matrix elements and security parameters for Quantum Key Distribution (QKD). In addition to putting past experiments on rigorous footing, we also develop a physical noise model and propose a novel setup that can further enhance the noise resistance of free-space quantum communication. 
\end{abstract}

\maketitle

\section{Introduction \label{sec:Intro}} 
Quantum entanglement is the defining feature of quantum mechanics. It serves as the main resource for the envisioned ``quantum internet'' \cite{Kimble2008}, which enables novel communication protocols such as super-dense coding \cite{denseCoding1992}, quantum teleportation \cite{teleportation1993}, and, notably, Quantum Key Distribution (QKD) \cite{Bennett_Brassard_1984, Ekert_1991}. 
QKD allows two remote parties to establish secret keys, even in the presence of an eavesdropper having access to unlimited computational resources. 
The keys obtained then can be used in any broader cryptographic setting, such as encrypting secret messages. 
The main technical challenge in building a robust large-scale quantum network is the unavoidable exponential signal loss introduced by optical fibers \cite{Pirandola2017}.
In contrast to classical information, unknown quantum states cannot be copied \cite{WootersZurek_1982}; therefore, signal amplification by means of repeaters is impossible for quantum signals.
One could resort to building quantum repeaters based on entanglement swapping, however, existing repeater designs are very far from being practical~\cite{Azuma_2023}.
An alternative approach to overcome the exponential loss in optical fibers is satellite-based Quantum Key Distribution. This method allows the connection of two distant communicating sites on Earth through links with substantially reduced loss, benefiting from signal loss scaling only quadratically with the distance.
Unfortunately, satellite-earth links are inherently vulnerable to outside noise, entering the measurement devices from the channel, hence basically limited to nighttime operations (see Refs. \cite{Han_22, Abasifard_2023, Krzic_2023, Li_2023, Bouchard_2021, Avesani_2021, Liao_2017} for recent advances to extend the operation time towards daylight operations by mainly experimental adaptations). This severely limits the practicality of such systems. 
In particular, for long-distance satellite links, this presents a heavy reduction of potential links and up-times. High-Dimensional (HD) entanglement \cite{Kwiat_1997, Barreiro_2005, Sheridan_2010, Martin_2017, Islam_2017, Bouchard_2018, Avesani_2021, Bulla_2022, sulimany2023} has proven to enhance background noise resistance \cite{Ecker_2019} in entanglement-distribution tasks, with HD entanglement in the time-domain \cite{Cozzolino_2019} being its most promising implementation for free-space applications. Previous work has successfully demonstrated the feasibility and advantages of HD entanglement in entanglement distribution and QKD \cite{Doda_2021, Bulla_2022, Bulla_2023}, but analyses have been rather heuristic and have relied on assumptions on the source-state that are not compatible with the security requirements of QKD. 
In this work, we provide a rigorous and clean theoretical analysis of a high-dimensional interferometry setup and show how it can be used to harness entanglement in the polarization-time domain for QKD. We develop a realistic noise model and propose an altered scheme that is experimentally simpler and removes assumptions on the entangled state present in previous work. An additional benefit of the proposed setup is an improved noise tolerance by a factor of almost two. This theoretically shifts operation hours from dawn to daytime without relying on substantial technological innovations, marking an important step towards full daytime operations.

\begin{figure*}
\includegraphics[width=1\textwidth]{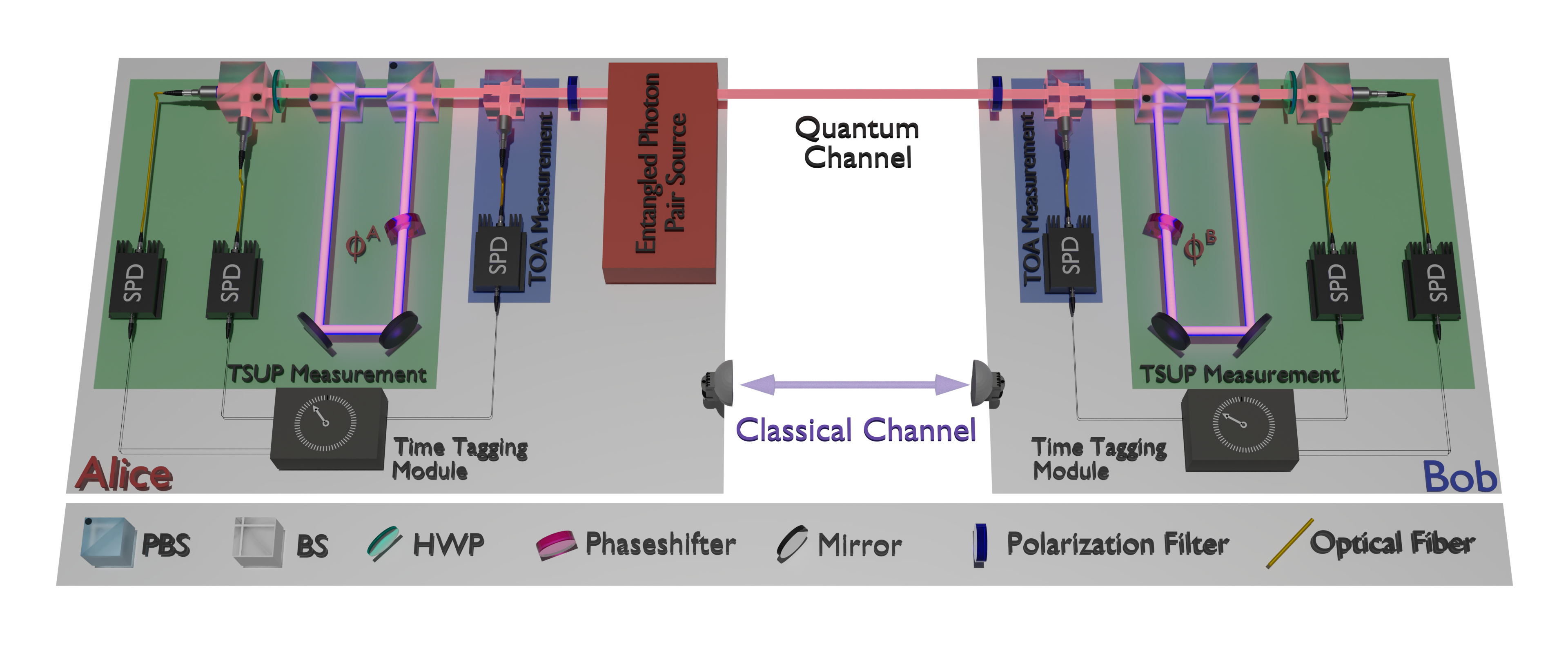}
\caption{Sketch of the setup analyzed. A photon source prepares entangled photons and sends them to two identical labs. Each of the labs is equipped with a polarizing filter which can be either inserted (Protocol 1) or removed (Protocol 2). Then an $\eta:1-\eta$ beamsplitter allows a choice between two measurements, followed by a time-resolving photon detector, which allows measuring the arrival time of incoming photons (TOA). In the second measurement setup, two polarizing beamsplitters (PBSs) and two mirrors allow the superposition of neighboring time-bins. Finally, a half-wave plate (HWP) that can be either inserted or removed together with another PBS and two time-resolving photon detectors (SPD) allow us to perform Time-Superposition (TSUP) measurements. Furthermore, both labs are connected via a classical channel, allowing  authenticated classical messages to be exchanged.\label{fig:Sketch_general} }
\end{figure*}
\noindent

\section{Setup description \& analysis \label{sec:Setting} }
The HD-QKD setup analyzed is built upon two identical measurement devices, which are placed in the communicating parties' labs, as well as an entangled photon pair source, which can be placed either in one of the parties' labs or in the middle and does not need to be trusted. For the present work, we assume the photon source is being placed in Alice's lab. However, we want to stress that our method applies to both scenarios.

We define the parties' time-reference points via synchronized coincidence windows which will henceforth be called 'time-frames'. If the source is placed in one of the parties' labs, the delay in arrival time in comparison to the lab of the second party is accounted for, i.e., the time-reference points of detection are brought into agreement.
The source produces general polarized, time-entangled photon pairs in $\left(\mathcal{H}_{\textrm{Pol}} \otimes \mathcal{H}_{\textrm{Time}}\right)^{\otimes 2}$,
$\ket{\Psi_{\textrm{target}}^{\textrm{ideal}}}\!:=\!\sum_{p_A,p_B \in \{\!H,V\!\}}\!\int_0^{\infty}\!\Psi(p_A,p_B,t) \ket{p_A,p_B,t,t}\!\! ~\textrm{d}\mu(t),
$
where $\mu(t)$ is some appropriate measure. Here, $\mathcal{H}_{\textrm{Pol}}$ is a two-dimensional Hilbert space representing the polarization degree of freedom, while, in principle, we require an infinite-dimensional Hilbert space $\mathcal{H}_{\textrm{Time}}$ to represent the temporal degree of freedom. However, any realistic time-resolving photon measurement discretizes the temporal degree of freedom into detection events of finitely many time-bins.\\ 
Let us call $T>0$, chosen such that $\forall p_A,p_B \in \{\text{H},\text{V}\}: \Psi(p_A,p_B,t)$ is almost constant in $[t_0,t_0+T]$ for some $t_0$, time-frame length and let $d$ be the number of time-bins. Consequently, the length of a single time-bin is given by $t_B := \frac{T}{d}$. Hence, effectively, we require a $d$-dimensional Hilbert space $\mathcal{H}_T$ to capture the temporal aspects of the photons under consideration. Then, our effective target state reads
\begin{equation}\label{eq:diskTargetState}
    \ket{\Psi_{\textrm{target}}^{\textrm{eff}}} := \sum_{p_A,p_B \in \{\text{H},\text{V}\} }\!\!\!\!\!\!\!\!c_{p_A,p_B}  \ket{p_A,p_B} \otimes \frac{1}{\sqrt{d}} \sum_{k=0}^{d-1} \ket{kk},
\end{equation}
where $c_{p_A,p_B} \in \mathbb{C}$ with $\sum_{p_A,p_B} |c_{p_A,p_B}|^2 = 1$. The shares of the photon pair are transmitted to Alice and Bob, respectively.

Each of their labs is equipped with a measurement setup described in Figure \ref{fig:Sketch_general}, where incoming photons pass an $\eta:1-\eta$ beamsplitter, which, with probability $\eta$, allows them to measure the Time-of-Arrival (TOA), and with probability $1-\eta$ the Temporal-Superposition (TSUP) of neighboring time-bins. We note that the same measurements can also be realized with active choice instead of the beamsplitter, such that both measurements are performed with only two detectors. For simplicity, we discuss the passive choice setup, however, the following analysis is also valid for active basis choice. The first arm of this setup consists of a simple time-resolving photon detector to capture the arrival time of incoming photons. In the second arm, two polarizing beamsplitters (PBSs) and two mirrors build a Franson interferometer~\cite{Franson_1989}, allowing to superpose two (not necessarily neighboring) time-bins. Finally, a half-wave plate (HWP) that can be either inserted or removed, together with another PBS and two time-resolving photon detectors (SPD) perform TSUP measurements. 

 For what follows, we adjust the long arm of the Franson interferometer such that it causes a delay by one time-bin, i.e., vertically polarized light with time-stamp $i-1$ meets horizontally polarized light with time-stamp $i$. While a possible basis for $\mathcal{H}_{\mathrm{Pol}}$ is given by $\{\ket{\text{H}}, \ket{\text{V}}\}$, consisting of vectors corresponding to horizontally and vertically polarized photons, a basis of $\mathcal{H}_{\mathrm{T}}$ is given by the ``time-bin states'' $\{\ket{n}\}_{n=0}^{d-1}$. This allows us to define the time-shift operation by its action on basis states, $\hat{T}: \mathcal{H}_T \rightarrow \mathcal{H}_T,~ \ket{n} \mapsto \ket{n+1}$ and the phase-shift operator $\hat{Q}_{\phi}: \mathcal{H}_{\textrm{Pol}} \otimes \mathcal{H}_T \rightarrow \mathcal{H}_{\textrm{Pol}} \otimes \mathcal{H}_T,~ \ket{p,n} \rightarrow e^{i\phi} \ket{p,n}$. Having introduced this notation, we can start to describe the action of the measurement setup. Let $\rho_{AB} \in \mathcal{D}\left(\mathcal{H}^A \otimes \mathcal{H}^B \right)$ be the joint quantum state that enters Alice's and Bob's lab. The action of the TOA measurement setup on this state is straightforward, as a detection event at time $i$ simply corresponds to a projection of the incoming state onto $\ket{i}$. Thus, the corresponding measurement operator for Alice and Bob respectively reads
\begin{equation}
    M^{A/B}(i) = \mathbbm{1}_{\textrm{Pol}} \otimes \ketbra{i}.
\end{equation}
It remains to find the corresponding operators for TSUP measurements. We show in Appendix \ref{apdx:AnalysisSetup} that the unitary representing the action of the TSUP measurement setup is given by
\begin{align}
\begin{aligned}
    \hat{U} :=& \ket{\text{HH}}\!\!\bra{\text{HH}} \otimes \mathbbm{1}_T\otimes\mathbbm{1}_T + \ket{\text{HV}}\!\!\bra{\text{HV}} \otimes \mathbbm{1}_T\otimes \hat{Q}_{\phi}\hat{T} \\
    + &\ket{\text{VH}}\!\!\bra{\text{VH}} \otimes \hat{Q}_{\phi}\hat{T}\otimes\mathbbm{1}_T + \ket{\text{VV}}\!\!\bra{\text{VV}} \otimes \hat{Q}_{\phi}\hat{T}\otimes\hat{Q}_{\phi}\hat{T}.
\end{aligned}
\end{align}
Note that the interferometer is followed by an half-wave plate that rotates the plane of polarization by $\frac{\pi}{4}$, mapping $\ket{\text{H}}$ to $\ket{\text{D}}$ and $\ket{\text{V}}$ to $\ket{\text{A}}$. This finally allows us to find the `effective measurement' performed on the input state $\rho_{AB}$, i.e., if $M_k$ denotes the measurement performed on the state $\rho_{\mathrm{out}}$ after passing the TSUP setup (but before the detectors), we aim for $\tilde{M}_k:= \ketbra{\tilde{\Psi}_k}$ such that 
\begin{align}
    \Tr{\rho_{AB} \tilde{M}_k } =& \Tr{\rho_{\textrm{out}} M_k} = \Tr{\hat{U}\rho_{AB}\hat{U}^{\dagger} M_k} \\
    =& \Tr{\rho_{AB}\hat{U}^{\dagger} M_k \hat{U}}.
\end{align}

 Let $a,b \in \{1,2\}$ label Alice's and Bob's detectors and denote by $i$ and $j$ their respective time-stamps. Then, we obtain the effective measurement operators $\tilde{M}_{a,b}(i,j,\phi^A, \phi^B) = \ketbra{\tilde{\Psi}_{a,b}(i,j,\phi^A,\phi^B)}$,
\begin{equation}\label{eq:DD_op}
\begin{aligned}
    \ket{\tilde{\Psi}_{1,1}(i,j,\phi^A, \phi^B)}:= &\hat{U}^{\dagger} \ket{\text{DD}, i,j} \\
    =& \ket{\tilde{\Psi}_1(i,\phi^A)} \otimes \ket{\tilde{\Psi}_1(j,\phi^B)}, 
\end{aligned}
\end{equation}

\begin{equation}\label{eq:DA_op}
\begin{aligned}
    \ket{\tilde{\Psi}_{1,2}(i,j,\phi^A, \phi^B)} := &\hat{U}^{\dagger} \ket{\text{DA}, i,j} \\ =&\ket{\tilde{\Psi}_1(i,\phi^A)} \otimes \ket{\tilde{\Psi}_2(j,\phi^B)},
\end{aligned}
\end{equation}

\begin{equation}\label{eq:AD_op}
\begin{aligned}
    \ket{\tilde{\Psi}_{2,1}(i,j,\phi^A, \phi^B)} := &\hat{U}^{\dagger} \ket{\text{AD}, i,j} \\
    =& \ket{\tilde{\Psi}_2(i,\phi^A)} \otimes \ket{\tilde{\Psi}_1(j,\phi^B)} ,
\end{aligned}
\end{equation}

\begin{equation}\label{eq:AA_op}
\begin{aligned}
    \ket{\tilde{\Psi}_{2,2}(i,j,\phi^A, \phi^B)} :=& \hat{U}^{\dagger} \ket{\text{AA}, i,j} \\
    =&\ket{\tilde{\Psi}_2(i,\phi^A)} \otimes \ket{\tilde{\Psi}_2(j,\phi^B)},
\end{aligned}
\end{equation}

where we defined 
\begin{equation}
 \begin{aligned}
    &\ket{\tilde{\Psi}_{x}(i,\phi)} :=\frac{1}{\sqrt{2}} \left( \ket{\text{H},i}+ (-1)^{x-1} e^{-i \phi}\ket{\text{V}, i-1} \right), \label{eq:PsiPM}
\end{aligned}   
\end{equation}
for $x \in \{1,2\}$. To ease notation, we define
$\Tilde{M}^A_{a}(i, \phi^A) := \ketbra{\tilde{\Psi}_a(i,\phi^A)}$ and $\Tilde{M}^B_{b}(j,\phi^B):=\ketbra{\tilde{\Psi}_b(j,\phi^B)}$, for $a,b \in\{1,2\}$ indicating which detector clicks and $i,j \in \{0,...,d-1\}$ marking the time-stamps on each side. These measurements with time-stamp $i$ correspond to positive operator-valued measure (POVM) elements associated with the detection time of a photon emitted at time $t_i$ that traveled the short interferometer path, or, equivalently, with the detection time of a photon emitted at time $t_i-1$ that traveled the long interferometer path.

Over the course of their experiment, Alice and Bob each measure either in the TOA or in the TSUP setting, recording clicks with time-stamps that are stored in different coincidence-click matrices. Let $a,b \in \{1,2\}$ label Alice's and Bob's detectors and denote by $i$ and $j$ their respective time-stamps. Depending on which measurement (TOA or TSUP) they chose, we obtain four possible measurement combinations that are stored in four different kinds of coincidence-click matrices. Note that we only need to label which detector clicked if a TSUP measurement was performed, as clicks in TOA do not discriminate polarization. In what follows, we use the following notation. If both measured the time of arrival, the corresponding coincidence-click element is $\mathrm{TT}(i,j)$ and if both measured temporal superposition, the corresponding coincidence-click element is given by $\mathrm{SS}_{a,b}(i,j)$, where $a$ and $b$ indicate which detector clicked. Since Alice and Bob each have two detectors for that case, we obtain four $\mathrm{SS}$ coincidence-click matrices in total. We follow the same naming convention for those rounds where Alice and Bob chose different measurements: if Alice measured TOA while Bob measured TSUP, we denote the corresponding coincidence-click element by $\mathrm{TS}_b(i,j)$, while for the opposite case where Alice measured TSUP and Bob TOA it reads $\mathrm{ST}_a(i,j)$, with two matrices for each of the cases. Based on the performed measurements, the coincidence-click matrix elements obtained by the present setup read
\begin{align}
 \textrm{TT}(i,j) &:= \Tr{\rho_{AB}\left(\mathbbm{1}_{\textrm{Pol}}^{\otimes 2} \otimes \ketbra{i,j}\right) } \label{eq:genCC} \\
    \textrm{SS}_{a,b}(i,j, \phi^A, \phi^B) &:= \Tr{\rho_{AB} \tilde{M}_{a,b}(i,j,\phi^A,\phi^B) }  \label{eq:genDDab}\\
    \textrm{TS}_{b}(i,j, \phi^B) &:= \Tr{\rho_{AB} \left(\left( \mathbbm{1}_{\textrm{Pol}} \otimes \ketbra{i} \right)\otimes \tilde{M}_b\left(j,\phi^B \right) \right) } \label{eq:genCDb}\\
     \textrm{ST}_{a}(i,j, \phi^A) &:= \Tr{\rho_{AB} \left( \tilde{M}_a(i,\phi^A) \otimes \left(\mathbbm{1}_{\textrm{Pol}} \otimes \ketbra{j}\right)  \right) }. \label{eq:genDCa}
\end{align}

\section{Problem and solution} \label{sec:ProbAndSol} 
In general, Alice and Bob obtain a quantum state $\rho \in \mathcal{D}\left(  \left(\mathcal{H}_{\textrm{Pol}}\otimes \mathcal{H}_{T}\right)^{\otimes 2}\right)$. Both record time-stamped clicks and correlate the ones within the temporal margin of the time-frame as coincidence-click matrices
$\textrm{CC}(i,j):=\#\textit{Clicks per frame in time-bin $i$ and $j$}$ which leads to the four different kinds of coincidence-click matrices given in Eqs. (\ref{eq:genCC}) - (\ref{eq:genDCa}).

While we can choose the state prepared by the source, the state Alice and Bob receive (or only Bob receives, in case the source is located in Alice's lab) is unknown, as in QKD, the channel connecting Alice and Bob is assumed to be fully under the control of an eavesdropper, called Eve. By performing measurements, Alice and Bob aim to certify that the temporal part of their shared state is entangled and decoupled from Eve's part.

In general, the interpretation of the measurements and their meaning for the time part of the density matrix depends on the polarization degree of freedom of the state Alice and Bob receive. A common approach so far has been to \textit{assume} (a) that temporal and polarization degrees of freedom are actually independent of each other, $\rho = \rho_{\textrm{Pol}}\otimes \rho_T$, or (b) that the polarization degree of freedom does not change while the photons travel through the quantum channel. The assumed perfect knowledge of the polarization degree of freedom has allowed us to directly interpret the measurements' effect on the time part of the density matrix. Alternatively, one could lift assumption (b) by performing state tomography of the polarization density matrix by adding and using an additional measurement arm at the cost of losing a certain fraction of the signals and making the analysis significantly more complicated.

In what follows, we propose a solution that simultaneously removes both assumptions (a) and (b) while, as a beneficial side-effect, improving the noise resistance of the setup and easing the practical complexity of the experiment significantly. As shown in Figure \ref{fig:Sketch_general}, we suggest adding an additional polarization filter set to let $D$-polarized photons pass at the entrance of both Alice's and Bob's lab before the signal meets the first beamsplitter and tune the source to produce $D$-polarized photon pairs. Thus, compared to earlier setups \cite{Bulla_2023}, we require only one time-resolving photon detector in the TOA measurement (and do not need an additional measurement arm to perform state tomography, as required under (b)). While this simple modification eases the experimental setup considerably, by forcing Alice's and Bob's joint state to be $\ketbra{DD}\otimes \rho_T$ after the polarization filters, we are able to go without any assumptions about the internal structure of the state that enters the lab and without requiring that only the time part is manipulated while passing the quantum channel. Furthermore, we expect our proposed protocol to be more favorable in the finite-size regime as we need to account for finite-size effects for fewer measurements. 

\section{Quantum Key Distribution} \label{sec:QKD}

Having clarified the setup (see Figure \ref{fig:Sketch_general}), let us now detail the proposed High-Dimensional (HD) Discrete-Variable (DV) Quantum Key Distribution (QKD) protocol. Therefore, Alice and Bob execute the following protocol.
\begin{enumerate}
    \item[1)] \textbf{\textit{State Preparation---}} A source generates photon pairs entangled in polarization and time 
     \begin{equation}\label{eq:targetStateP1}
    \ket{\Psi_{\textrm{target}}^{\textrm{P1}}} :=  \ket{\text{DD}} \otimes \frac{1}{\sqrt{d}} \sum_{k=0}^{d-1} \ket{kk},
\end{equation}
    and sends them to Alice and Bob.
   
    \item[2)] \textbf{\textit{Measurement---}} Alice and Bob each perform either a TOA or a TSUP measurement, depending on independent random bits $\mathbb{P}_{A/B} \in \{0,1\}$. This step can be implemented passively via a beamsplitter.
\end{enumerate}

Steps 1) and 2) are repeated $N$-times, where $N$ is assumed to be very large.

\begin{enumerate}
    \item[3)] \textbf{\textit{Announcement \& Sifting---}} Alice and Bob publicly announce their measurement choices for every round via the classical channel and sift rounds where they have performed different measurements.
    \item[4)] \textbf{\textit{Parameter Estimation \& Key Generation---}} Next, they disclose some of their results from measurements of each basis to perform statistical tests. If the tests are passed, they use the TOA measurements to create a common raw key by performing a key map, where logical bit-values are assigned to the measurement results. Otherwise, they abort the protocol and start again at step 1). 
    \item[5)] \textbf{\textit{Error-correction \& Privacy Amplification---}} By means of classical algorithms Alice and Bob reconcile their raw keys and perform privacy amplification to eliminate the potential eavesdropper's knowledge about the key.
\end{enumerate}
We take our proposed protocol (`Protocol 1') as an example to illustrate our method and compare it to a protocol used in earlier work \cite{Bulla_2022, Bulla_2023}, which relied on assumptions (a) and (b) discussed in the previous section. For this protocol (`Protocol 2'), the source is set to prepare 
\begin{equation}\label{eq:targetStateP2}
    \ket{\Psi_{\textrm{target}}^{\textrm{P2}}} :=  \frac{\ket{\text{HH}}+\ket{\text{VV}}}{\sqrt{2}} \otimes \frac{1}{\sqrt{d}} \sum_{k=0}^{d-1} \ket{kk},
\end{equation}
and the polarization filters at the entrance of both Alice's and Bob's labs are removed. Note that the rates reported for this protocol are only upper bounds, while the key rates we report for Protocol 1 are reliable lower bounds.

Let us start with the discussion of Protocol 1. By construction, independently of what happens to the quantum signal while traveling through the quantum channel, Alice's and Bob's joint quantum state $\rho_{AB}$ reads $\ketbra{\text{DD}} \otimes \rho_T$, where $\rho_T$ is an arbitrary quantum state in $\mathcal{D}(\mathcal{H}_T^{\otimes 2})$. We apply Eqs. (\ref{eq:genCC}) - (\ref{eq:genDCa}) to relate coincidence-click matrix elements with density matrix elements. For the TOA measurements, it follows directly that the coincidence-clicks correspond to the diagonal entries of the time-density matrix,
\begin{equation}
    \textrm{TT}(i,j) = \bra{i,j|\rho_T}\ket{i,j}.
\end{equation}

For the TSUP measurements, using the $D/A$ representation from Eq. (\ref{eq:PsiPM}), we obtain
\begin{align}
    \textrm{SS}_{1,1}(i,j) &= \frac{1}{4} \bra{i+,j+|\rho_T}\ket{i+,j+}, \label{eq:genDD11}\\
    \textrm{SS}_{1,2}(i,j) &= \frac{1}{4} \bra{i+,j-|\rho_T}\ket{i+,j-},\label{eq:genDD12}\\
    \textrm{SS}_{2,1}(i,j) &= \frac{1}{4} \bra{i-,j+|\rho_T}\ket{i-,j+},\label{eq:genDD21}\\
    \textrm{SS}_{2,2}(i,j) &= \frac{1}{4} \bra{i-,j-|\rho_T}\ket{i-,j-} \label{eq:genDD22},
\end{align}
where $i,j \in \{1,2,\dotsc, d-1\}$ and $\ket{i\pm} := \frac{\ket{i}\pm \ket{i-1}}{\sqrt{2}}$, i.e., we set $\phi^A = \phi^B = 0$. To ease notation, we omitted the arguments $\phi^A$ and $\phi^B$. Note that these click events have a direct physical interpretation. If on both sides detector $1$ clicks, this means both have measured a positive phase between two neighboring time-bins labeled by $i$ and $j$, while clicks of detector $2$ on both sides indicate a negative phase. If opposite detectors click, this indicates that they have measured different phases between neighboring time-bins.

Finally, the mismatched measurements yield
\begin{align}
    \textrm{TS}_{1}(i,j) &= \frac{1}{2} \bra{i,j+|\rho_T}\ket{i,j+}, \label{eq:genCD1}\\
    \textrm{TS}_{2}(i,j) &= \frac{1}{2} \bra{i,j-|\rho_T}\ket{i,j-},\label{eq:genCD2}\\
    \textrm{ST}_{1}(i,j) &= \frac{1}{2} \bra{i+,j|\rho_T}\ket{i+,j},\label{eq:genDC1}\\
    \textrm{ST}_{2}(i,j) &= \frac{1}{2} \bra{i-,j|\rho_T}\ket{i-,j} \label{eq:genDC2}.
\end{align}

For ease of presentation, we assume now that $d$ is even in what follows, in order to explain how these click matrices can be interpreted. We want to emphasize that our method is not limited to this case and that this choice is solely made for illustration purposes and generalizes straight-forwardly to arbitrary dimensions.

Note that the right-hand sides of the click equations are composed of matrix elements of the time-density matrix. It can be seen directly that the TOA clicks ($\textrm{TT}$) already correspond to a POVM,
\begin{equation}\label{eq:P1_POVM1}
  \mathcal{M}_0^{P1} :=  \left\{ \ketbra{i,j} \right\}_{i,j=0}^{d-1},
\end{equation}
giving rise to a basis $\mathcal{B}_0 := \mathcal{B}_0^{A} \otimes \mathcal{B}_0^B$, where $\mathcal{B}_0^{A/B} := \left\{\ket{i}\right\}_{i=0}^{d-1}$ spans single time-bin subspaces. The corresponding `basis-click matrix' reads
\begin{equation}
    C_{\mathcal{M}_0^{P1}}(i,j) := \textrm{TT}(i,j),
\end{equation}
where we use the natural order $\{\ket{0}, \ket{1}, \hdots, \ket{d-1} \}$.

In contrast, the TSUP measurements can even be used to construct two POVM measurements. For the first of the two corresponding bases, $\mathcal{B}_1 := \mathcal{B}_1^{A} \otimes \mathcal{B}_1^B$, notice that $\ket{i\pm}$ for fixed $i$ spans two-dimensional time-bin subspaces. Thus, we first consider odd $i$ and obtain $\mathcal{B}_1^{A/B} := \left\{ \ket{(2k-1)\pm} \right\}_{k=1}^{\frac{d}{2}}$ as a basis for Alice's, respectively Bob's, $d$-dimensional temporal Hilbert space. Consequently, a short calculation shows that
\begin{align}
\begin{aligned}
      \mathcal{M}_1^{P1} := &\left\{ \ketbra{i+,j+}, \ketbra{i+,j-}, \right.\\
      &~\left.\ketbra{i-,j+}, \ketbra{i-,j-} \right\}_{i,j \textrm{ odd} }  
\end{aligned}
\end{align}
forms another POVM, induced by the bases $\mathcal{B}_1^{A/B}$ on Alice's and Bob's side, respectively. The corresponding click-matrix reads
\begin{widetext}
\begin{align}
    \begin{aligned}
    &C_{\mathcal{M}_1^{P1}} :=4 \left(\begin{smallmatrix}
        \textrm{SS}_{1,1}(1,1) & \textrm{SS}_{1,2}(1,1) & \textrm{SS}_{1,1}(1,3) & \textrm{SS}_{1,2}(1,3) & \hdots & \textrm{SS}_{1,1}(1,d-1)  & \textrm{SS}_{1,2}(1,d-1)\\
        \textrm{SS}_{2,1}(1,1) & \textrm{SS}_{2,2}(1,1) & \textrm{SS}_{2,1}(1,3) & \textrm{SS}_{2,2}(1,3) & \hdots & \textrm{SS}_{2,1}(1,d-1) & \textrm{SS}_{2,2}(1,d-1)\\
        \textrm{SS}_{1,1}(3,1) & \textrm{SS}_{1,2}(3,1) & \textrm{SS}_{1,1}(3,3) & \textrm{SS}_{1,2}(3,3) & \hdots &\textrm{SS}_{1,1}(3,d-1) & \textrm{SS}_{1,2}(3,d-1) \\
        \textrm{SS}_{2,1}(3,1) & \textrm{SS}_{2,2}(3,1) & \textrm{SS}_{2,1}(3,3) & \textrm{SS}_{2,2}(3,3) & \hdots & \textrm{SS}_{2,1}(3,d-1) & \textrm{SS}_{2,2}(3,d-1)\\
        \vdots & \vdots & \vdots & \vdots & \ddots & \vdots & \vdots \\
         \textrm{SS}_{1,1}(d-1,1) & \textrm{SS}_{1,2}(d-1,1) & \textrm{SS}_{1,1}(d-1,3) & \textrm{SS}_{1,2}(d-1,3) & \hdots & \textrm{SS}_{1,1}(d-1,d-1)  & \textrm{SS}_{1,2}(d-1,d-1)\\
        \textrm{SS}_{2,1}(d-1,1) & \textrm{SS}_{2,2}(d-1,1) & \textrm{SS}_{2,1}(d-1,3) & \textrm{SS}_{2,2}(d-1,3) & \hdots & \textrm{SS}_{2,1}(d-1,d-1) & \textrm{SS}_{2,2}(d-1,d-1)\\
    \end{smallmatrix}\right),
     \end{aligned}
\end{align}
\end{widetext}
where we have ordered the basis vectors $\{\ket{1+}, \ket{1-}, \ket{3+}, \ket{3-}, \hdots, \ket{(d-1)+}, \ket{(d-1)-}\}$.

 Finally, for the last basis, $\mathcal{B}_2 := \mathcal{B}_2^{A} \otimes \mathcal{B}_2^B$, we consider even $i$. Based on our measurement setup we are not able to directly measure elements spanning the boundary subspaces $\textrm{span}\{\ket{0}\}$ and $\textrm{span}\{\ket{d-1}\}$ in the TSUP basis with even $i$ and hence need to substitute them by combinations of elements where the parties have used different measurements (so, the $\textrm{TS}$- and $\textrm{ST}$-clicks). We obtain $\mathcal{B}_2^{A/B} := \{\ket{(2k)\pm}\}_{k=1}^{\frac{d}{2}-1} \cup \left\{\ket{0}, \ket{d-1}\right\}$. This translation of projective measurements resulting in coincidence-click matrices to POVM elements on only the temporal Hilbert space allows us now to normalize the click matrices correctly. We obtain for the third POVM
 
\begin{align}
\begin{aligned}
       \mathcal{M}_2^{P1} &:= \left\{ \ketbra{i\pm,j\pm} \right\}_{i,j>0, \atop{\textrm{even}}} \cup \left\{ \ketbra{i, j\pm} \right\}_{{i \in \{0, d-1\}}\atop{j>0, \textrm{ even }} }\\
       &\cup \left\{ \ketbra{i\pm, j} \right\}_{{i>0, \textrm{ even }}\atop{j \in \{0, d-1\}} } \cup \left\{ \ketbra{i,j}\right\}_{i,j \in \{0, d-1\}}. 
\end{aligned}
\end{align}    
The corresponding click-matrix reads 
\begin{widetext}
\begin{align}
\begin{aligned}
   &C_{\mathcal{M}_2^{P1}} :=
   &\begin{psmallmatrix}
        \textrm{TT}(0,0) & 2\textrm{TS}_{1}(0,2) & 2\textrm{TS}_{2}(0,2) &      2\textrm{TS}_{1}(0,4) &        2\textrm{TS}_{2}(0,4) &        
        \hdots & 2\textrm{TS}_{2}(0,d-2) & \textrm{TT}(0,d-1)
        \\        2\textrm{ST}_{1}(2,0) & 4\textrm{SS}_{1,1}(2,2) & 4\textrm{SS}_{1,2}(2,2) &        4\textrm{SS}_{1,1}(2,4) &         4\textrm{SS}_{1,2}(2,4) & 
        \hdots & 4\textrm{SS}_{1,2}(2,d-2) &2\textrm{ST}_{1}(2,d-1) \\
        2\textrm{ST}_{2}(2,0) & 4\textrm{SS}_{2,1}(2,2) & 4\textrm{SS}_{2,2}(2,2) &         4\textrm{SS}_{2,1}(2,4) &         4\textrm{SS}_{2,2}(2,4) & 
        \hdots & 4\textrm{SS}_{2,2}(2,d-2) & 2\textrm{ST}_{2}(2,d-1) \\
        2\textrm{ST}_{1}(4,0) & 4\textrm{SS}_{1,1}(4,2) & 4\textrm{SS}_{1,2}(4,2) &         4\textrm{SS}_{1,1}(4,4) &         4\textrm{SS}_{1,2}(4,4) & 
        \hdots & 4\textrm{SS}_{1,1}(4,d-2) & 2\textrm{ST}_{1}(4,d-1) \\   
        2\textrm{ST}_{2}(4,0) & 4\textrm{SS}_{2,1}(4,2) & 4\textrm{SS}_{2,2}(4,2) &         4\textrm{SS}_{2,1}(4,4) &         4\textrm{SS}_{2,2}(4,4) & 
        \hdots & 4\textrm{SS}_{2,2}(4,d-2) & 2\textrm{ST}_{2}(4,d-1) \\        
        \vdots & \vdots & \vdots & \vdots & \vdots & \ddots & \vdots \\
        2\textrm{ST}_{2}(d-2,0) & 4\textrm{SS}_{2,1}(d-2,2) & 4\textrm{SS}_{2,2}(d-2,2) & 
        4\textrm{SS}_{2,1}(d-2,4) & 
        4\textrm{SS}_{2,2}(d-2,4) & 
        \hdots & 2\textrm{SS}_{2,2}(d-2,d-2) &2\textrm{ST}_{2}(d-2,d-1)\\
        \textrm{TT}(d-1,0) & 2\textrm{TS}_{1}(d-1,2) & 2\textrm{TS}_{2}(d-1,2) & 
        2\textrm{TS}_{1}(d-1,4) & 
        2\textrm{TS}_{2}(d-1,4) & 
        \hdots & 2\textrm{TS}_{2}(d-1,d-2)&\textrm{TT}(d-1,d-1)
    \end{psmallmatrix},
\end{aligned}
\end{align}
\end{widetext}
where the order of the basis vectors is $\{ \ket{0}, \ket{2+} \ket{2-}, \hdots, \ket{(d-2)+}, \ket{(d-2)-},\ket{d-1} \}$. Note that the different weight factors ensure the correct normalization of the click-matrix. Finally, we want to emphasize the importance of the overlap between the subspaces spanned by $\mathcal{B}_1$ and $\mathcal{B}_2$, which allows us to certify high-dimensional entanglement.

For comparison, we conduct a similar analysis for Protocol 2 under the assumption that the polarization degree of freedom is known. Therefore, the rates reported for this protocol represent only upper bounds on the secure key rate. For details, we refer the reader to Appendix \ref{APDX:HHVVProtocol}.

\section{Noise model \label{sec:NoiseModel}}
Entangled photons are produced by the source and then travel through a (free-space) quantum channel before entering imperfect detection devices, where they cause clicks. In this section, we physically model how various physical processes influence the coincidence-click matrices used to calculate key rates for our setup. Similarly to Refs. \cite{Doda_2021} and \cite{Chapman_2022}, besides loss --- i.e., the process that a photon coming from the source is lost on its way to the detectors --- we identify two main origins of noise, i.e., processes that add photons, hence detector clicks. While photons are traveling through the free-space channel, some of the photons may scatter on molecules present in the direct line between the source and the detector. Thus, the probability of losing a photon increases with increasing distance between sender and receiver and is given by a probability of photon loss $P_{\mathrm{loss}}^A$ (for the channel between Alice and the source) and $P_{\mathrm{loss}}^B$ (for the channel between Bob and the source). 
Due to imperfections in the detection process, not all incoming photons cause a detector click. We measure the click probability for a given photon by a number $\eta_D \in [0,1]$, called the detection efficiency. For simplicity we assume that the detector efficiency is the same for all the detectors in our setup.

Next, let us turn to processes that insert photons to our setup that do not originate from the source. The main source of this type of noise is environmental photons (like those coming from the Sun) being detected. Besides that, detector imperfections sometimes cause clicks even if no photons are present. Such clicks are called dark counts and we measure this effect as dark count rate (in dark counts per second). 

We provide a detailed derivation of our noise model in Appendix \ref{apdx:NoiseModel} and give here only the key ideas. Due to the independent nature of all processes described, it is reasonable to model both the photon pair production and the dark counts as well as environmental photons as Poisson-distributed quantities. The pair production rate, the dark count rate, etc. are then simply given by the expectation of the corresponding Poisson-distributed random variable. Again, motivated by the independent nature of all sources of noise, we aim to quantify the influence of noise by one single parameter, the isotropic noise parameter such that
\begin{equation}
    \rho(v) = v \ketbra{\Psi_{\textrm{target}}} + (1-v) \frac{1}{d^2} \mathbbm{1}_{d^2 \times d^2},
\end{equation}
where $v = \frac{P_\textrm{Good}}{ P_{\textrm{CC}}(1,1)}$. Here, $P_\textrm{Good}$ denotes the probability that a coincidence-click is caused by a photon pair originating from the source, while $P_{\textrm{CC}}(1,1)$ is the probability for one coincidence-click.
In Appendix \ref{apdx:NoiseModel}, we derive expressions for $P_\textrm{Good}$ and $P_{\textrm{CC}}(1,1)$ for both protocols. In what follows, we assume that the source is placed in Alice's lab, which is in line with the practical implementation of entanglement-based QKD setups, such as the free-space link between Vienna and Bisamberg \cite{Bulla_2022, Bulla_2023}. Thus, Alice's share of the entangled photon pair is not subject to channel loss ($P_{\mathrm{loss}}^A = 0$) and does not experience noise due to environmental photons. 

Note that we do not consider detector jitter in our noise model. This is an effect, in which the time of arrival of photons is mislabeled due to the clock imprecision. One would expect that the effect of detector jitter on the observed key rate of protocols is non-negligible, especially for higher values of $d$, which are obtained by making the time-bins rather short. While this is true for the time frame and time-bin definition presented in this manuscript, in practical implementations, one can mitigate this effect by using non-neighboring time-bins to define the time-frames, as discussed in \cite{Bulla_2023}. In particular, a $d$-dimensional time-frame is defined as a collection of $d$ non-neighboring time-bins that are separated by time intervals equal to the interferometer delay, which is typically much larger than the length of a time-bin. 
Defining the frames in this way makes mislabelling due to jitter very likely to produce two single-click events in non-matched time-frames, which are then discarded as single-click events. This reduction in error, however comes at the cost of reduced coincidence rate, therefore optimal key rate per coincidence and optimal key rate per second are typically not achieved for the same local dimension.
Finally, defining time-frames in this way does not affect the analysis of the protocol and we have opted for a more traditional time-frames definition in this manuscript for simplicity.\\

\section{Numerical Method} \label{sec:NumericalMethod}
The asymptotic secure key rate $R^{\infty}$ of a QKD protocol is lower-bounded by the Devetak-Winter bound \cite{Devetak_Winter_2006}, which reads 
\begin{equation}
 R^{\infty} \geq S(A|E)-H(A|B),   
\end{equation}
where $S(A|E)$ is the conditional von Neumann entropy of Alice's key register given Eve's quantum register, quantifying the amount of information not known by Eve, and $H(A|B)$ is the Shannon entropy between Alice's and Bob's raw key, representing the amount of information leaked during the error-correction phase of the protocol. While the latter quantity is accessible from the observed statistics, to obtain the first, we have to minimize over all quantum states $\rho_{ABE}$ compatible with Alice's and Bob's observations. Therefore, we use a method developed in Ref. \cite{Araujo_2022} that exploits a semi-definite program (SDP) hierarchy converging to the first term in the Devetak-Winter formula. The first term can be rewritten by means of the quantum relative entropy $D(\rho||\sigma) := \Tr{\rho \log(\rho)}-\Tr{\rho \log(\sigma)}$ for density matrices $\rho$ and $\sigma$ and reads
\begin{equation}
    S(A|E) = - D\left( \rho_{\tilde{A}E} \right|\!\left| \mathbbm{1}_A \otimes \rho_E \right),
\end{equation}
where $\rho_{\tilde{A}E} := \sum_a \ketbra{a} \otimes \mathrm{Tr}_{AB}\left[(\ketbra{a}\otimes\mathbbm{1}_{BE})\rho_{ABE}\right]$. Following Ref. \cite{Brown_2023} one then can obtain a convergent sequence of semi-definite programs for the quantum relative entropy using Gauß-Radau quadrature. We implement this numerical method and solve semi-definite programs (SDP) with Gauß-Radau parameter $m = 10$, to find lower bounds on the secure key rate for protocol 1 (and upper bounds for protocol 2). Experimental observations then serve as constraints for this semi-definite program. Hence, the optimization for our particular problem reads
\begin{widetext}
  \begin{align}
    \begin{aligned}
        \min_{\sigma, \{\zeta_i^a, \eta_i^a, \theta_i^a\}_{a,i}}& c_m + \sum_{i=1}^{m}\sum_{a=0}^{d-1} \frac{w_i}{t_i \log(2)} \Tr{(\ketbra{a} \otimes\mathbbm{1}_B) \left(\zeta_i^a+{\zeta_i^a}^{\dagger}+(1-t_i)\eta_i^a \right) + t_i \theta_i^a}\\
        \text{ s.t. }& \\
        & \Tr{\sigma} = 1,\\
        &\forall a, i:~ \Gamma_{a,i}^1:=\begin{pmatrix}
            \sigma & \zeta_i^a\\
            {\zeta_i^a}^{\dagger} & \eta_i^a
        \end{pmatrix} \geq 0, \\
        &\forall a, i:~ \Gamma_{a,i}^2:=\begin{pmatrix}
            \sigma & {\zeta_i^a}^{\dagger} \\
            \zeta_i^a& \theta_i^a
        \end{pmatrix} \geq 0, \\
        & \forall k:~ \Tr{E_k \sigma} = f_k,
    \end{aligned}
\end{align}
\end{widetext}

where we defined $\Theta(M) := \mathrm{Tr}_E\left[ \rho_{ABE}(\mathbbm{1}_{AB}\otimes M_E^{\top}) \right]$ and set\newpage
\begin{align}
 \sigma &:= \Theta(\mathbbm{1})\\
 \zeta_i^a &:= \Theta({Z_i^a})\\
 \eta_i^a &:= \Theta({Z_i^a}^{\dagger}Z_i^a) \\
 \theta^a &:= \Theta(Z_i^a{Z_i^a}^{\dagger})
\end{align}
and the $Z_i^a$ are arbitrary complex matrices. Furthermore $c_m := \sum_{i=1}^{m} \frac{w_i}{t_i \log(2)}$, where the $w_i>0$, $\sum_i w_i=1$, $w_m=\frac{1}{m^2}$ are Gauß-Radau weights and $t_i \in (0,1]$, $t_m = 1$. For more details we refer interested readers to Ref. \cite{Araujo_2022}.

The constraints of the form $\Tr{E_k \sigma} = f_k$ are due to our experimental observations, i.e., the operators $E_k$ are chosen from $\{ \mathcal{M}_0^{P_z}, \mathcal{M}_1^{P_z}, \mathcal{M}_2^{P_z}\}$ and the corresponding scalar right-hand sides $f_k$ are given by the matrices $C_{\mathcal{M}_0^{P_z}}, C_{\mathcal{M}_1^{P_z}}$ and $C_{\mathcal{M}_2^{P_z}}$, where $z\in \{1,2\}$ selects between the two protocols we consider (see Section \ref{sec:QKD} and Appendix \ref{APDX:HHVVProtocol}).

\section{Results}\label{sec:Results}

We illustrate our method under the realistic noise model we derived (see Section \ref{sec:NoiseModel} and Appendix \ref{apdx:NoiseModel}), taking photon losses, background noise induced by solar photons, detector inefficiencies, and dark counts into account. Recall that in Protocol 1, the source prepares the state 
\begin{equation}
    \ket{\Psi_1} = \ket{\text{DD}}\otimes \frac{1}{\sqrt{d}}\sum_{k=0}^{d-1}\ket{kk},
\end{equation}
and Alice and Bob each add a polarization filter, which is set to $\text{D}$, right after the photon enters their respective lab, such that they are aligned on polarization of the target state. Consequently, as outlined earlier, the state after the polarization filter has the form $\rho_{AB} = \ketbra{\text{DD}}\otimes \rho_T$. We wish to highlight that neither the form of the polarization part nor the tensor product structure are mere assumptions, as they are enforced by the filter. This allows a direct and clean analysis of the QKD setup without any additional, unjustified assumptions. For comparison, we also consider a second protocol, which was already discussed in earlier works \cite{Bulla_2022, Bulla_2023}. There, the source produces the target state
\begin{align}\label{eq:entPol}
    \ket{\Psi_2} = \frac{\ket{\text{HH}}+\ket{\text{VV}}}{\sqrt{2}} \otimes \frac{1}{\sqrt{d}}\sum_{k=0}^{d-1}\ket{kk}.
\end{align}
\begin{figure}
\includegraphics[width=0.48\textwidth]{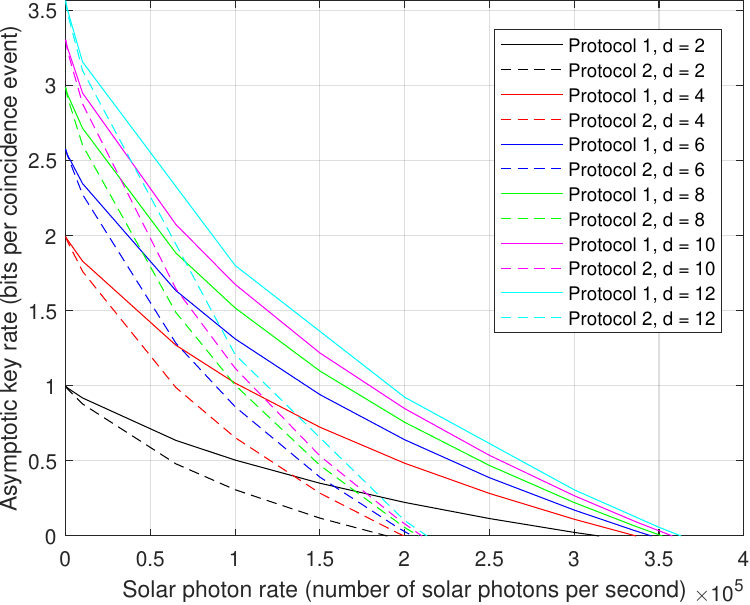}
\caption{Lower bounds on the secure key rate for Protocol 1 (solid) and upper bounds on the secure key rate for Protocol 2 (dashed) for various dimensions as a function of the solar photon rate. We set the time-frame length to $T= 5.4\times 10^{-9} s$ and the probability of channel loss to $99.7\%$, which corresponds to $25.2$~dB loss. Furthermore, we assume a detection efficiency of $90\%$ as well as a dark count rate of $100/s$. Finally, we set the pair production rate to $0.1/T$, which corresponds to approximately $18.5 \times 10^6$ photons per second.
\label{fig:plot_for_paper} }
\end{figure}   

The original motivation to consider a QKD protocol where a state with maximally entangled polarization degree of freedom (DOF) \eqref{eq:entPol} is distributed, was to implement a postselection free Franson interferometer. In other words, in case of noiseless polarization propagation through the channel, Alice's and Bob's photons would always take the same path (long-long or short-short) through the Franson interferometer, thus decreasing the number of coincidences that are not directly useful to witness entanglement.
Such a setup, however, leads to some experimental restrictions. First of all, one cannot put any polarization filter in the entry of Alice's respectively Bob's lab. In addition, one also needs to enforce certain restrictions on the capability of the adversary, which is undesirable in adversarial scenarios. Namely, either one has to assume that (a) the polarization remains unchanged over the channel (which is a very strong assumption as one would expect noise to primarily affect the polarization-degree of freedom) and still forms a tensor product with the time part,
\begin{equation}\label{eq:stateAssump}
    \rho_{AB} =\\ 
    \tfrac{\ketbra{\text{HH}} + \ket{\text{HH}}\!\!\bra{\text{VV}} + \ket{\text{VV}}\!\!\bra{\text{HH}} +  \ketbra{\text{VV}}}{2} \otimes \rho_T,
\end{equation}
or (b) the received state has at least tensor-product structure, $\rho_{AB} = \rho_{\textrm{Pol}} \otimes \rho_T$ (also an unjustified assumption), where $\rho_{\textrm{Pol}}$ might have changed over the channel, and needs to be determined by performing additional tomography in the polarization degree of freedom. Assumption (b) is technically weaker than assumption (a), but, it is also much harder to analyze because a change in the polarization would lead to different overall measurements implemented by the Franson interferometer. In this case one would hope that the implemented tomography reveals a high fidelity of the polarization DOF to the maximally entangled state so that the postselection free property would still be present and contribute to the overall key rate. 
Since this assumption was used in the existing literature \cite{Bulla_2022}, we decided to use it for comparison. 
We refer to a protocol with entangled polarization and assumption (a) as Protocol 2 and label Protocol 1 as our new results with polarization prepared in the $\ket{\text{DD}}$ state and polarization filters in both measurement apparatuses. We emphasize that this leads to a comparison of a lower bound on the secure key rate of Protocol 1 with an upper bound for Protocol 2.
 Besides allowing a clean and rigorous analysis without any assumptions, we intuitively expect the additional polarization filter to increase the resistance against solar photons, which are the main source of noise in free-space and satellite QKD applications, as half of the unpolarized solar photons are blocked, while (in the ideal case) none or (in reality) only a small fraction of the source photons are. 

For our simulations, we used specific parameter values in accordance with the Free-Space Link experiment between Vienna and Bisamberg \cite{Bulla_2022, Bulla_2023}. 
The length of a time-frame was set to $T_0 := 5.4 \times 10^{-9}s$, the dark count rate to $100/s$, the detection efficiency to $\eta_D=90\%$, the channel loss to $P_{\textrm{loss}} = 99.7\%$, corresponding to a loss of $25.2 dB$, and the pair production rate to $0.1$ per time-frame, corresponding to roughly $18.5$~MHz repetition rate. 
Further, consistent with the experiments that inspired our work, we assume that the photon source is located in Alice's lab. In Figure \ref{fig:plot_for_paper}, we present secure key rates for Protocol 1 and compare them to upper bounds on the secure key rate for Protocol 2 for time dimensions of $d=2,4,6,8,10$ and $12$. This means we keep the time-frame size $T$ fixed while changing the number of time-bins $d$. It is common to plot secure key rates over the error rate which is related to the visibility via $1-v-\frac{1-v}{d}$. However, depending on the particular noise model, the visibility may be a function of the dimension and protocol specifics and, therefore, is a suboptimal measure to compare different dimensions. Thus, we have chosen to compare the protocols and various dimensions as a function of the solar photon rate, which, in our opinion, provides a fair and practical comparison.

\begin{figure}
\includegraphics[width=0.49\textwidth]{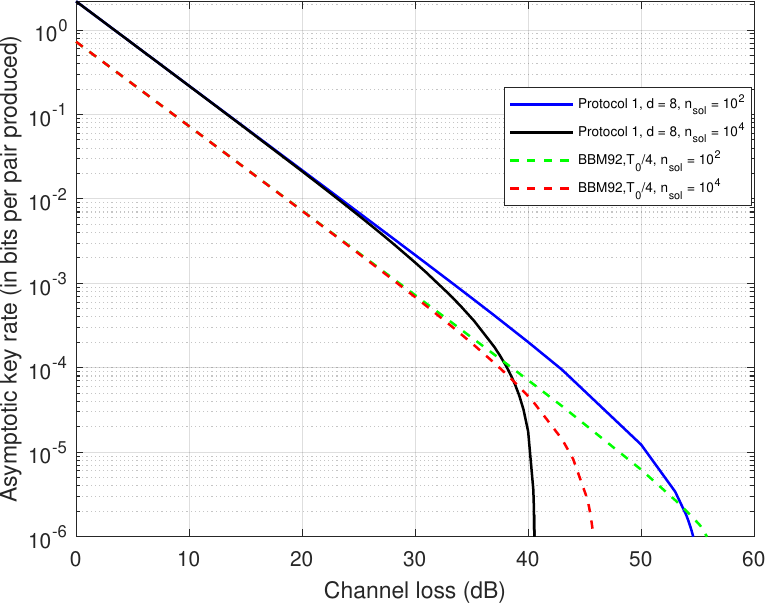}
\caption{Comparison of Protocol 1 (solid lines) for $d=8$ with the corresponding key rates obtained for BBM92 (dashed lines) for solar photon rates of $n_{\text{sol}} \in \{10^2, 10^4\}$. 
\label{fig:ComparisonPlot}}
\end{figure}

We note that one should not directly read the values for different dimensions $d$ as the optimal performance for QKD implementations at different dimensions in general. For such a comparison, one would need to optimize the photon pair production and the time-frame size for each value of $d$ for the largest achievable key rate per second, which is dependent on many parameters beyond the scope of this work.
In Figure \ref{fig:ComparisonPlot}, we compare the performance of our protocol with BBM92 regarding loss. Therefore, we keep all system parameters the same as in Figure \ref{fig:plot_for_paper}, but vary the photon loss probability and plot curves for different solar photon rates. Again, we employ the noise model presented in Appendix \ref{apdx:NoiseModel} to relate physical parameters to click matrices and we derive the comparison key rates for the BBM92 protocol using the fact that our Protocol~$1$ reduces to the BBM92 protocol if two time-bins are used in a single frame. Technically, this case is a special case of our protocol, as we envision choosing the optimal dimension dynamically depending on background noise and loss. To keep the comparison meaningful, we always compare a situation where the photon pair production rate, the solar photon rate, and the clock precision (bin size) are the same and vary the probability of loss. This means we compare Protocol~$1$ with frame-size fixed to $T=5.4\times 10^{-9}$ and local dimension $d$ to a BBM92 protocol with the same bin size and local dimension $2$ (i.e., the frame size of the BBM92 protocol is equal to $T/(d/2)$. This comparison is meaningful, as we assume that the source brightness is at a point that saturates the detectors on one side of a free-space link and cannot be increased anymore. In Figure \ref{fig:ComparisonPlot} we present key rate curves comparing Protocol $1$ for $d=8$ with BBM92 for solar photon rate $n_{\text{sol}} \in \{10^2, 10^4\}$. One can observe that our protocol outperforms BBM92 in lower loss regions up to approximately $39$dB. 
In particular, our protocol outperforms BBM92 also when the loss is approximately $25$dB, (which is consistent with the experiment reported in Ref. \cite{Bulla_2022} and the loss in LEO satellite scenarios \cite{Gruneisen2021}) and the background solar photon rate is set to $10^{4}$ per second, which is roughly consistent with the light conditions at sunrise (see Ref. \cite{Bulla_2022}). However, we also see that the region where Protocol~$1$ outperforms BBM92 shrinks with increasing background photon rate.

Lastly, we note that we assume ideal one-way error correction~\cite{Brassard_ECC} and we leave performance optimizations and examination of alternative reconciliation routines~\cite{Cascade_1994,Tupkary_2023,Mink2014} for future work.
Our new protocol (Protocol 1) consistently exhibits significantly higher key rates, even compared to the upper bound for the existing protocol (Protocol 2), especially as the number of solar photons increases. Additionally, Protocol 1 demonstrates tolerance to nearly twice as many solar photons per second as Protocol 2. It is crucial to reiterate that the curves presented for Protocol 2 rely on unjustified assumptions (see our earlier discussion), serving as mere upper bounds, while the analysis of Protocol 1 avoids these assumptions, offering reliable lower bounds, which is the main point of our paper.

\section{Discussion \label{sec:Discussion}}
Previous works have successfully demonstrated the distribution and certification of high-dimensional entanglement. While certain assumptions are suitable for these scenarios, in order to utilize high-dimensional entanglement for quantum key distribution, one requires an assumption-free analysis of the measurement setups used. In this work, we discuss assumptions present in previous works and suggest a simple modification of the measurement setup that helps to remove those assumptions. Our main contribution is a clean and rigorous analysis of a modified measurement setup, suitable not only for high-dimensional entanglement distribution but also for high-dimensional quantum key distribution. We also develop a realistic noise model for free-space QKD links, taking external factors such as solar light, atmospheric channel loss, and imperfections of the measurement devices, such as dark counts and detector inefficiencies, into account. Then, we apply a numerical security proof method to calculate lower bounds on the asymptotic secure key rate under the noise model developed. 
For comparison, we also consider a previously used measurement setup, that relies on certain assumptions that are not compatible with the requirements of QKD. Therefore, rates reported for that setup are only upper bounds. Nevertheless, our analysis shows that the rigorous lower bounds obtained for our proposed setup outperform the upper bounds obtained for the previous setup relying on unjustified assumptions both in terms of key rate and noise tolerance. Additionally, our modification simplifies the experimental setup significantly.  We also compare our proposed protocol to BBM92 and observe better performance w.r.t. loss for relevant loss regimes.

Given that the unmodified setup (within the frame of mentioned assumptions) was able to transmit key during early daytime in summer \cite{Bulla_2022} in urban atmospheric conditions, this seemingly modest increase of a factor of almost 2 in solar photons is actually a significant advance towards a full-day operation. Technical improvements such as sharper frequency filters and adaptive optics to correct for atmospheric turbulences are expected to improve key rate and noise tolerance further (see, for example, the experimental improvements in Refs. \cite{Han_22, Abasifard_2023, Krzic_2023, Li_2023, Bouchard_2021, Avesani_2021, Liao_2017}). While the key rates presented are asymptotic, we anticipate that the advantages of our new protocol will become even more pronounced in the finite-size regime due to our simplified setup, which necessitates fewer measurements. The experimental simplicity of our proposal (only one interferometer needs to be stabilized), together with the results of our analysis, guides a practical path toward full-day satellite QKD. Besides a detailed finite-size analysis, interesting future directions include finding rapidly computable quantifiers instead of numerical SDPs, required to employ online adaptive subspace postselection, which can increase noise tolerance further \cite{Doda_2021,Hu_2021}.

\begin{acknowledgements}
We thank Mateus Ara\'{u}jo for numerous discussions about SDP implementations.

This work has received funding from the Horizon-Europe research and innovation programme under grant agreement No 101070168 (HyperSpace).

\centering
\includegraphics[width=0.25\textwidth]{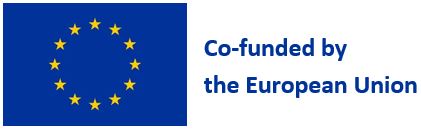} 
\end{acknowledgements}

\bibliography{Bibliography}

\appendix
\onecolumngrid
\section{Analysis of the Franson Setup}\label{apdx:AnalysisSetup} 
In what follows, we derive the measurement operators for Time-Superposition measurements in more detail than in the main part.

Denote by $x$ and $y$ the horizontal and vertical input ports of a polarizing beamsplitter, respectively, and by $x'$ and $y'$ the corresponding outputs. Note that in our setup, the first beamsplitter has only one input port and the second beamsplitter has only one output port. Then, the action of a PBS is given by the unitary
\begin{equation}
    \hat{U}_{\textrm{PBS}} = \left( \ket{\text{H}}\!\!\bra{\text{H}} \otimes\mathbbm{1}_T \otimes  \ket{x'}\!\!\bra{x} + \ket{\text{H}}\!\!\bra{\text{V}} \otimes \mathbbm{1}_T \otimes \ket{x'}\!\!\bra{y} + \ket{\text{V}}\!\!\bra{\text{H}} \otimes \mathbbm{1}_T \otimes \ket{y'}\!\!\bra{x} + \ket{\text{V}}\!\!\bra{\text{V}} \otimes \mathbbm{1}_T \otimes \ket{y'}\!\!\bra{y}\right).
\end{equation}
According to the description of our measurement setup above, only those parts in the upper $y'$-part experience a time as well as a phase shift. Thus, let
\begin{align}
    \hat{T}_y &:= \mathbbm{1}_{\textrm{Pol}} \otimes \mathbbm{1}_T \otimes \ket{x'}\!\!\bra{x'} +  \mathbbm{1}_{\textrm{Pol}} \otimes \hat{T} \otimes \ket{y'}\!\!\bra{y'}\\
    \hat{Q}_y &:= \mathbbm{1}_{\textrm{Pol}} \otimes \mathbbm{1}_T \otimes \ket{x'}\!\!\bra{x'} + \hat{Q}_{\phi} \otimes \ket{y'}\!\!\bra{y'}
\end{align}
be the operators describing these shifts. They allow us to describe the action of the whole setup by applying them one after another in the correct order
\begin{align}
    \hat{U}_{I} := \hat{U}_{\textrm{PBS}} \hat{Q}_y \hat{T}_y \hat{U}_{\textrm{PBS}}.
\end{align}
As the output of the first PBS is directed to the input of the second PBS, we simply write $(x',y')$ for the input ports of the second PBS, while we proceed with $(x'', y'')$ for its outputs. We obtain
\begin{align}
    \hat{U}_I = \left(\ket{\text{H}}\!\!\bra{\text{H}} \otimes \mathbbm{1}_T + \ket{\text{V}}\!\!\bra{\text{V}} \otimes \hat{Q}_{\phi}\hat{T} \right) \otimes \ket{x''}\!\!\bra{x} + \left(\ket{\text{H}}\!\!\bra{\text{H}} \otimes \hat{Q}_{\phi} \hat{T} + \ket{\text{V}}\!\!\bra{\text{V}} \otimes \mathbbm{1}_T \right) \otimes \ket{y''}\!\!\bra{y}.
\end{align}
Note that in our case, input $y$ is empty, so effectively, we only need to consider the first term -- horizontally polarized states pass the interferometer untouched while vertically polarized states experience a time as well as a phase shift. As we put our measurement devices in the output port $x''$ of the interferometer, we can drop the register $x''$ for ease of notation.

Finally, we obtain the unitary operator representing the action of the whole interferometer setup on Alice's and Bob's joint state $\rho_{AB}$,
\begin{align}
\begin{aligned}
    \hat{U} &:= \hat{U}_I \otimes \hat{U}_I\\
    &= \ket{\text{HH}}\!\!\bra{\text{HH}} \otimes \mathbbm{1}_T\otimes\mathbbm{1}_T + \ket{\text{HV}}\!\!\bra{\text{HV}} \otimes \mathbbm{1}_T\otimes \hat{Q}_{\phi}\hat{T} + \ket{\text{VH}}\!\!\bra{\text{VH}} \otimes \hat{Q}_{\phi}\hat{T}\otimes\mathbbm{1}_T + \ket{\text{VV}}\!\!\bra{\text{VV}} \otimes \hat{Q}_{\phi}\hat{T}\otimes\hat{Q}_{\phi}\hat{T}.
\end{aligned}
\end{align}

The interferometer is followed by a half-wave plate and another polarizing beamsplitter, followed by two detectors, one in each arm of the PBS such that the right detector measures horizontally polarized photons, while the upper detector measures vertically polarized photons. However, the half-wave plate rotates the plane of polarization as $U_{\textrm{HWP}} \ket{\text{H}} = \ket{\text{D}}$ and $U_{\textrm{HWP}} \ket{\text{V}} = \ket{\text{A}}$, i.e., it allows us to measure diagonally polarized photons in the right arm and antidiagonally polarized photons in the upper arm. The corresponding measurements project onto $\ket{\text{D}, i}$ and $\ket{\text{A}, i}$ respectively. 
This can be used to derive the effective measurements, as carried out in the main part of this manuscript.

\section{Analysis of Protocol 2}\label{APDX:HHVVProtocol}
Finally, we turn to the second protocol where the target state is $\ket{\Psi_{\textrm{target}}} = \frac{\ket{\text{HH}}+\ket{\text{VV}}}{\sqrt{2}} \otimes \frac{1}{\sqrt{d}} \sum_{k=0}^{d-1} \ket{kk}$. Unlike in Protocol 1, there is no polarization filter, so we have to \textit{assume} that the polarization remains unchanged when the state passes the quantum channel. Thus, Alice's and Bob's shared state reads $\rho_{AB} = \frac{\ket{\text{HH}}\!\bra{\text{HH}} + \ket{\text{HH}}\!\bra{\text{VV}}+\ket{\text{VV}}\!\bra{\text{HH}}+\ket{\text{VV}}\!\bra{\text{VV}}}{2} \otimes \rho_T$.
As for Protocol 1, it follows directly that the coincidence-clicks for the TOA measurements correspond to the diagonal entries of the time-density matrix,
\begin{equation}
    \textrm{TT}(i,j) = \bra{i,j|\rho_T}\ket{i,j}.
\end{equation}
Applying Eq. (\ref{eq:genDDab})  to $\rho_{AB}$ yields
\begin{align}
    \textrm{SS}_{1,1}(i,j) &= \frac{1}{8} \left(\bra{i+,j+|\rho_T}\ket{i+,j+} + \braket{i+,j+|\rho_T}{i-,j-} +  \braket{i-,j-|\rho_T}{i+,j+} + \braket{i-,j-|\rho_T}{i-,j-}\right) \\
    &=\textrm{SS}_{2,2}(i,j), \label{eq:P3:DD11}\\
    \textrm{SS}_{1,2}(i,j) &= \frac{1}{8} \left(\bra{i+,j-|\rho_T}\ket{i+,j-} + \braket{i+,j-|\rho_T}{i-,j+} +  \braket{i-,j+|\rho_T}{i+,j-} + \braket{i-,j+|\rho_T}{i-,j+}\right)\\
    & = \textrm{SS}_{2,1}(i,j).\label{eq:P3:DD12}
\end{align}
As $\textrm{SS}_{1,1}(i,j)$ and $\textrm{SS}_{2,2}(i,j)$ as well as $\textrm{SS}_{1,2}(i,j)$ and $\textrm{SS}_{2,1}(i,j)$ are equal, we can combine those elements respectively into 'same phase` and 'opposite phase` clicks,
\begin{align}
 \textrm{SS}_{s}(i,j) &:=\textrm{SS}_{1,1}(i,j)+\textrm{SS}_{2,2}(i,j)\label{eq:P3:DefDDs} \\
 \textrm{SS}_{o}(i,j) &:= \textrm{SS}_{1,2}(i,j)+\textrm{SS}_{2,1}(i,j).\label{eq:P3:DefDDo} 
\end{align}

For the mismatched measurements, we obtain from Eqs. (\ref{eq:genCDb})-(\ref{eq:genDCa})
\begin{align}
 \textrm{TS}_{1}(i,j) &= \frac{1}{2} \bra{i,j+|\rho_T}\ket{i,j+},\label{eq:P3:DefCD1} \\
 \textrm{TS}_{2}(i,j) &= \frac{1}{2} \bra{i,j-|\rho_T}\ket{i,j-},\label{eq:P3:DefCD2} \\ 
 \textrm{ST}_{1}(i,j) &= \frac{1}{2} \bra{i+,j|\rho_T}\ket{i+,j}, \label{eq:P3:DefDC1}\\
 \textrm{ST}_{2}(i,j) &= \frac{1}{2} \bra{i-,j|\rho_T}\ket{i-,j}.\label{eq:P3:DefDC2}
\end{align}

As for Protocol 1, one can see immediately that
\begin{equation}\label{eq:P2_POVM1}
  \mathcal{M}_0^{P1} :=  \left\{ \ketbra{i,j} \right\}_{i,j=0}^{d-1}
\end{equation}
forms a POVM induced by the bases $\mathbbm{B}_0^{A/B}$ on Alice's and Bob's side respectively with corresponding click-matrix
\begin{equation}
    C_{\mathcal{M}_0^{P1}}(i,j) := \textrm{TT}(i,j),
\end{equation}
where we have chosen the natural order $\{ \ket{0}, \ket{1}, ..., \ket{d-1}\}$. Next, we aim to find a second POVM, which requires some preparations. From the definitions made in Eqs. (\ref{eq:P3:DefDDs}) and (\ref{eq:P3:DefDDo}), it follows directly that
\begin{align}
    \textrm{SS}_{s}(i,j) &= \frac{1}{2} \Tr{\rho_T \ketbra{\Phi^{A,B}_s(i,j)}}\\
     \textrm{SS}_{o}(i,j) &= \frac{1}{2} \Tr{\rho_T \ketbra{\Phi^{A,B}_o(i,j)}},   
\end{align}
with
\begin{align}
   \ket{\Phi^{A,B}_s(i,j)} := \frac{1}{\sqrt{2}}\left( \ket{i+,j+}+\ket{i-,j-} \right), \\
   \ket{\Phi^{A,B}_o(i,j)} := \frac{1}{\sqrt{2}}\left( \ket{i+,j-}+\ket{i-,j+} \right).
\end{align}
A short calculation shows that
\begin{align*}
    \ketbra{\Phi^{A,B}_s(i,j)} + \ketbra{\Phi^{A,B}_o(i,j)} = \ketbra{i,j} + \ketbra{i-1,j-1}
\end{align*}
and
\begin{align*}
    \ketbra{i,j+} + \ketbra{i,j-} &= \ketbra{i,j} + \ketbra{i,j-1}\\
    \ketbra{i+,j} + \ketbra{i-,j} &= \ketbra{i,j} + \ketbra{i-1,j}.
\end{align*}
Next, we combine those measurement operators and obtain
\begin{align*}
 &\sum_{i,j=1}^{d-1} \left( \ketbra{\Phi^{A,B}_s(i,j)} + \ketbra{\Phi^{A,B}_o(i,j)}\right) \\
 +&\sum_{j=1}^{d-1} \left( \ketbra{0,j+}+ \ketbra{0,j-} +   \ketbra{d-1,j+}+ \ketbra{d-1,j-}\right) \\
 +& \sum_{i=1}^{d-1}  \left( \ketbra{i+,0}+ \ketbra{i-,0} +   \ketbra{i+,d-1}+ \ketbra{i-,d-1}\right)\\
 +& \left( \mathbbm{1}_{d^2\times d^2} - \ketbra{00} - \ketbra{d-1,d-1} \right)\\
 =& 3 \mathbbm{1}_{d^2\times d^2}.
\end{align*}

Thus, after scaling all elements by $\frac{1}{3}$, we obtain the second POVM,
\begin{equation}
    \begin{aligned}
        \mathcal{M}_1^{P2} := &\left\{\frac{1}{3}\left( \mathbbm{1}_{d^2\times d^2} - \ketbra{00} - \ketbra{d-1,d-1} \right), \frac{1}{3}\ketbra{\Phi^{A,B}_s(i,j)}, \frac{1}{3}\ketbra{\Phi^{A,B}_o(i,j)}, \right. \\
        &~~~\frac{1}{3}\ketbra{0,j+}, \frac{1}{3}\ketbra{0,j-}, \frac{1}{3}\ketbra{d-1,j+}, \frac{1}{3}\ketbra{d-1,j+},\\
        &~~\left. \frac{1}{3}\ketbra{i+,0}, \frac{1}{3}\ketbra{i-,0}, \frac{1}{3}\ketbra{i+,d-1}, \frac{1}{3}\ketbra{i-,d-1}   \right\}_{i,j=1}^{d-1}.
    \end{aligned}
\end{equation}
Consequently, taking both the renormalization by $\frac{1}{3}$ and Eqs. (\ref{eq:P3:DefDDs}) - (\ref{eq:P3:DefDC2}) into account, the corresponding clicks are normalized as follows,
\begin{equation}
    \begin{aligned}
        1 &= \frac{1}{3}\left(1- \textrm{TT}(0,0) - \textrm{TT}(d-1,d-1) \right) + \frac{2}{3} \sum_{i,j=1}^{d-1} \left(\textrm{SS}_s(i,j)+\textrm{SS}_o(i,j)\right)\\
        &+ \frac{2}{3} \sum_{j=1}^{d-1} \left(\textrm{TS}_1(0,j) +\textrm{TS}_2(0,j) + \textrm{TS}_1(d-1,j) + \textrm{TS}_2(d-1,j)\right) \\
        &+ \frac{2}{3} \sum_{i=1}^{d-1} \left(\textrm{ST}_1(i,0) +\textrm{ST}_2(i,0) + \textrm{ST}_1(i,d-1) + \textrm{ST}_2(i,d-1)\right).
    \end{aligned}
\end{equation}
As we have already used the clicks from the computational basis measurement ($\textrm{TT}$-clicks), we do not expect any contribution to the key rates from these clicks. Since they simply introduce redundant constraints into our semi-definite program, we can remove those elements/clicks when we formulate the SDP as long as we account for them during normalization.

In contrast to Protocol 1, we have already exhausted our measurements so that we cannot build a third POVM.

\section{Details regarding the noise model} \label{apdx:NoiseModel}

As already mentioned in the main text, we have to consider various origins of noise. In particular, there are noise effects due to the interaction of the photons with the environment, mainly photons coming from the Sun, and due to the imperfect channels and detectors, so we take into account four different kinds of imperfections: 
    \begin{itemize}
        \item[(1)] Channel loss: Photons might get lost on the way from the source to the labs. 
        \item[(2)] Detection inefficiency: Due to detection inefficiencies, incoming photons cause a click with probability $\eta_D \in [0,1]$.
        \item[(3)] Environmental photons: Photons coming from the residual environment (like those coming from the Sun) can cause clicks.
        \item[(4)] Dark counts: Imperfections of the detectors can cause clicks even in the absence of photons.
    \end{itemize}
    
Let $P_{\textrm{prod}}(n)$ be the probability distribution of $n$ polarized photon pairs produced by the source per time-frame, let $P_{\textrm{loss}}^A$ and $P_{\textrm{loss}}^B$ denote the probabilities for source photons to get lost on their way from the source to Alice's and Bob's lab, respectively, and let $\eta_A$ and $\eta_B$ be the detection efficiency parameters of Alice's and Bob's detectors respectively. The probability that the environment, including the Sun, produces $n$ (unpolarized) photons per time-frame on Alice's side is given by $P_{\textrm{env}}^A(n)$ and on Bob's side by $P_{\textrm{env}}^B(n)$. In case the incoming photons pass a polarization filter that is aligned with the polarization produced by the source (Protocol 1), on average, half of the (unpolarized) environmental photons are blocked. Moreover, for each of the detectors, $n$ dark counts per time-frame may occur with a probability of $P_{\textrm{dark}}(n)$. These considerations yield:
\begin{itemize}
    \item The probability that $n$ photon pairs are produced per time-frame $T$ is given by $P_{\textrm{prod}}(n)$.
    \item Some photons get lost, so the probability of having $b_1$ (respectively $b_2$) photons left for Alice and Bob after the lossy channel is given by
    \begin{align*}P_1^A(b_1) &= \sum_{i=b_1}^{\infty}P_{\textrm{prod}}(i) (1-P_{\textrm{loss}}^A)^{b_1} P_{\textrm{loss}}^{i-b_1} \binom{i}{b_1},\\
    P_1^B(b_2) &= \sum_{i=b_2}^{\infty}P_{\textrm{prod}}(i) (1-P_{\textrm{loss}}^B)^{b_2} P_{\textrm{loss}}^{i-b_2} \binom{i}{b_2}\end{align*}
    respectively. These remaining photons are still polarized.
    \item The probability of $q$ sunlight and further environmental photons being produced within one time-frame $T$ on Alice's and Bob's side is $P_{\textrm{env}}^A(q)$ and $P_{\textrm{env}}^B(q)$, respectively.
    \item When the photons pass the polarization filter (which is placed only in Protocol 1), photons coming from the source remain mainly untouched, while half of the photons stemming from the Sun get blocked. Therefore, the probabilities that $e_1$ (respectively $e_2$) environmental photons pass Alice's and Bob's PBS are given by 
    \begin{align*}
        P_{\textrm{env}}'^A(e_1) &= \sum_{j=e_1}^{\infty} P_{\textrm{env}}^A(j) \left(\frac{1}{2}\right)^{e_1}\ \left(1-\frac{1}{2}\right)^{j-e_1} \binom{j}{e_1}=\sum_{j=e_1}^{\infty}\binom{j}{e_1} \left(\frac{1}{2}\right)^j P_{\textrm{env}}^A(j),\\
        P_{\textrm{env}}'^B(e_2) &=\sum_{j=e_2}^{\infty}\binom{j}{e_2} \left(\frac{1}{2}\right)^j P_{\textrm{env}}^B(j)
    \end{align*}
    respectively.
    \item Now the environmental photons are combined with the photons coming from the source. The probability of having $f_1$ photons in Alice's lab after the polarization filter therefore is given by 
    \begin{align*}
        P_2^A(f_1) &= \sum_{l=0}^{f_1}P_1^A(l) P_{\textrm{env}}'^A(f_1 - l)\\
        &= \sum_{l=0}^{f_1}\sum_{i=l}^{\infty}P_{\textrm{prod}}(i) (1-P_{\textrm{loss}}^A)^l (P_{\textrm{loss}}^A)^{(i-l)} \binom{i}{l} \sum_{j=f_1-l}^{\infty}\binom{j}{f_1-l} \left(\frac{1}{2}\right)^j P_{\textrm{env}}^A(j)\\
        &= \sum_{l=0}^{f_1}\sum_{i=l}^{\infty}\sum_{j=f_1-l}^{\infty}\binom{i}{l} \binom{j}{f_1-l} P_{\textrm{prod}}(i) P_{\textrm{env}}^A(j) (1-P_{\textrm{loss}}^A)^l (P_{\textrm{loss}}^A)^{(i-l)} \left(\frac{1}{2}\right)^j
    \end{align*}
    and similarly for Bob the probability of having $f_2$ photons in his lab after the filter is 
    \begin{align*}
        P_2^B(f_2) &= \sum_{l=0}^{f_2}\sum_{i=l}^{\infty}\sum_{j=f_2-l}^{\infty}\binom{i}{l} \binom{j}{f_2-l} P_{\textrm{prod}}(i) P_{\textrm{env}}^B(j) (1-P_{\textrm{loss}}^B)^l (P_{\textrm{loss}}^B)^{(i-l)} \left(\frac{1}{2}\right)^j.
    \end{align*}
    
    Since the source produces photon pairs, the joint probability of Alice and Bob having $n_1$ and $n_2$ photons respectively after the filter is found to be
    \begin{align}\label{eq:P_formula}
    P(n_1, n_2) = \sum_{s=0}^{\infty}\sum_{m_1 =0}^{\min\{s,n_1\}}\sum_{m_2 = 0}^{\min\{s,n_2\}}  P_{\textrm{prod}}(s)  & (1-P_{\textrm{loss}}^A)^{m_1} (P_{\textrm{loss}}^A)^{(s-m_1)}\binom{s}{m_1}\\  \cdot & (1-P_{\textrm{loss}}^B)^{m_2} (P_{\textrm{loss}}^B)^{(s-m_2)}\binom{s}{m_2} P_{\textrm{env}}'^A(n_1-m_1) P_{\textrm{env}}'^B(n_2-m_2)\\
    =\sum_{s=0}^{\infty}\sum_{m_1 = 0}^{\min\{s,n_1\}}\sum_{m_2 = 0}^{\min\{s,n_2\}}\binom{s}{m_1} \binom{s}{m_2} & P_{\textrm{prod}}(s) (1-P_{\textrm{loss}}^A)^{m_1} (P_{\textrm{loss}}^A)^{(s-m_1)}  (1-P_{\textrm{loss}}^B)^{m_2} (P_{\textrm{loss}}^B)^{(s-m_2)}\\
    \cdot \sum_{j=n_1-m_1}^{\infty}&\binom{j}{n_1-m_1} \left(\frac{1}{2}\right)^j P_{\textrm{env}}^A(j)  \sum_{k=n_2-m_2}^{\infty}\binom{k}{n_2-m_2} \left(\frac{1}{2}\right)^k P_{\textrm{env}}^B(k)\\
    = \sum_{s=0}^{\infty}\sum_{m_1 = 0}^{\min\{s,n_1\}}\sum_{m_2 = 0}^{\min\{s,n_2\}}\sum_{j=n_1-m_1}^{\infty}\sum_{k=n_2-m_2}^{\infty}&\binom{s}{m_1} \binom{s}{m_2} \binom{j}{n_1-m_1} \binom{k}{n_2-m_2}  P_{\textrm{prod}}(s) P_{\textrm{env}}^A(j)  P_{\textrm{env}}^B(k)\\  
    \cdot (1-P_{\textrm{loss}}^A)^{m_1} (1-P_{\textrm{loss}}^B)^{m_2} &(P_{\textrm{loss}}^A)^{(s-m_1)}  (P_{\textrm{loss}}^B)^{(s-m_2)}(P_{\textrm{loss}}^A)^{(s-m_1)} (P_{\textrm{loss}}^B)^{(s-m_2)} \left(\frac{1}{2}\right)^{(j+k)}.
    \end{align}

     \item Next, we take dark counts and detector inefficiencies into account. As the protocol discards all the events where there is a multiclick or no click in a time-frame, we only have to examine the case with exactly one dark count and no genuine photon being detected or when there is no dark count, and exactly one photon causes a detector click. This yields the probabilities for Alice and Bob each having exactly one click per time-frame in the same time-bin when there are $n_1$ (source or environmental) photons left on Alice's side and $n_2$ (source or environmental) photons left on Bob's side, 
    \begin{align*}
        P(\text{click}|n_1 \text{ photons}) &= P_{dark}(1)  (1-\eta_A)^{n_1} + P_{dark}(0) (1-\eta_A)^{(n_1-1)} \eta_A  \binom{n_1}{1}\\
        P(\text{click}|n_2 \text{ photons}) &= P_{dark}(1)  (1-\eta_B)^{n_2} + P_{dark}(0) (1-\eta_B)^{(n_2-1)} \eta_B \binom{n_2}{1}.
    \end{align*}
    \item Finally, the probability of exactly one coincidence click per time-frame in the same time-bin can be calculated as 
    \begin{align*}
        P_{TT}(1,1) := \sum_{n_1 = 0}^{\infty}\sum_{n_2 = 0}^{\infty}P(\text{click}|n_1 \text{ photons})  P(\text{click}|n_2 \text{ photons})  P(n_1, n_2).
    \end{align*}
\end{itemize}

All photon contributions mentioned are independent of each other and also photons from each of the origins mentioned are produced independently of other photons of the same origin. Hence, one can model these influences by Poisson distributions: 
\begin{align*}
    &P_{\textrm{prod}}(n) = \frac{(\lambda_p  T)^n  e^{-\lambda_p(T)}}{n!} =: \frac{C_p^n  e^{-C_p}}{n!}, \qquad P_{\text{dark}}(n) = \frac{(\lambda_d T)^n  e^{-\lambda_d(T)}}{n!} =: \frac{C_d^n  e^{-C_d}}{n!}
\end{align*}
\begin{align*}&P_{\textrm{env}}^A(n) = \frac{(\lambda_e^A  T)^n  e^{-\lambda_e^A(T)}}{n!} =: \frac{(C_{e}^A)^n  e^{-C_{e}^A}}{n!}, \qquad P_{\textrm{env}}^B(n) = \frac{(\lambda_e^B  T)^n  e^{-\lambda_e^B(T)}}{n!} =: \frac{(C_{e}^B)^n e^{-C_{e}^B}}{n!}
\end{align*}
Using the distribution of $P_{\textrm{env}}^{A/B}(n)$, we see that $P_{\textrm{env}}'^{A/B}(n)$ is distributed according to  \begin{align*}
    P_{\textrm{env}}'^{A/B}(n) &=\sum_{k=n}^{\infty}\binom{k}{n} \left(\frac{1}{2}\right)^k \frac{(C_{e}^{A/B})^k e^{-C_{e}^{A/B}}}{k!} = e^{-C_{e,A/B}} \sum_{k=n}^{\infty} \frac{k!}{n!(k-n)!}\frac{(C_{e,A/B})^k}{k! \cdot 2^k}\\
    & = \frac{e^{-C_{e}^{A/B}}}{n!} \sum_{k=n}^{\infty}\frac{(C_{e}^{A/B})^k}{2^k (k-n)!} \overset{j:=k-n}{=} \frac{e^{-C_{e,A/B}}}{n!} \sum_{j=0}^{\infty}\frac{(C_{e}^{A/B})^{j+n}}{2^{j+n} \cdot j!} = \frac{e^{-C_{e}^{A/B}} (C_{e}^{A/B})^{n}}{n! \cdot 2^n} \sum_{j=0}^{\infty}\frac{(C_{e}^{A/B})^j}{2^j \cdot j!}\\
    & = \frac{e^{-C_{e}^{A/B}}}{n!} \left(\frac{C_{e}^{A/B}}{2}\right)^n e^{\frac{C_{e}^{A/B}}{2}} = \frac{e^{\frac{-C_{e}^{A/B}}{2}}}{n!} \left(\frac{C_{e}^{A/B}}{2}\right)^n .
\end{align*}

Plugging $C_e$ into (\ref{eq:P_formula}), we obtain
\begin{align} \label{eq:P_formula_new}
    P(n_1, n_2) = \sum_{s=0}^{\infty}\sum_{m_1 = 0}^{\min\{s,n_1\}}\sum_{m_2 = 0}^{\min\{s,n_2\}} \sum_{j=n_1-m_1}^{\infty}\sum_{k=n_2-m_2}^{\infty}\binom{s}{m_1} \binom{s}{m_2}  &\binom{j}{n_1-m_1} \binom{k}{n_2-m_2} P_{\textrm{prod}}(s)   \frac{\left(C_{e}^{A}\right)^j  e^{-C_{e}^{A}}}{j!} \\  \frac{\left(C_{e}^{B}\right)^k  e^{-C_{e}^{B}}}{k!}  (1-P_{\textrm{loss}}^A)^{m_1} (1-P_{\textrm{loss}}^B)^{m_2} (P_{\textrm{loss}}^A)^{(s-m_1)}  &(P_{\textrm{loss}}^B)^{(s-m_2)} \left(\frac{1}{2}\right)^{(j+k)}.
\end{align}

With the formulas for $P_{\textrm{env}}'^{A/B}$, this yields 
\begin{align}
\begin{split}
\label{eq:P_n1_n2}
     P(n_1, n_2) = e^{-\frac{(C_{e}^{A}+C_{e}^{B})}{2}}\cdot \sum_{s=0}^{\infty}P_{\textrm{prod}}(s)\cdot (s!)^2\cdot  \sum_{m_1=0}^{\min\{s,n_1\}} \frac{(1-P_{\textrm{loss}}^A)^{m_1} (P_{\textrm{loss}}^A)^{s-m_1} (\frac{C_{e}^{A}}{2})^{n_1-m_1}}{m_1! (s-m_1)! (n_1-m_1)!}\\
     \qquad \cdot  \sum_{m_2=0}^{\min\{s,n_2\}} \frac{(1-P_{\textrm{loss}}^B)^{m_2} (P_{\textrm{loss}}^B)^{s-m_2} (\frac{C_{e}^{B}}{2})^{n_2-m_2}}{m_2! (s-m_2)! (n_2-m_2)!}.
\end{split}
\end{align}

To ease notation, we define
\begin{align*} I^{A/B}(s,n_j):=\sum_{m_i=0}^{\min\{s,n_j\}} \frac{(1-P_{\textrm{loss}}^{A/B})^{m_j}  (P_{\textrm{loss}}^{A/B})^{s-m_j} (\frac{C_{e}^{A/B}}{2})^{n_j-m_j}}{m_j! (s-m_j)! (n_j-m_j)!}, \qquad j \in \{1,2\}.
\end{align*}
For practical reasons, we choose $C_{p}$ such that the expected number of photon pairs per time-frame is much lower than $1$ so that the probability of more than one photon pair is close to zero, $P_{\textrm{prod}}(s>1) \approx 0$. We use this assumption, that it is extremely rare that more than one photon pair is emitted by the source in one time-frame and therefore can be neglected, also in our code for the calculation of the key rates.\\
Then, we obtain
\begin{align*}
    I^{A/B}(0,n) &= \frac{(C_{e}^{A/B})^{n}}{n!}, \\
    I^{A/B}(1,n) &= \begin{cases}
        P_{\textrm{loss}}^{A/B}, n=0, \\
        \frac{P_{\textrm{loss}}^{A/B} \left(\frac{C_{e}^{A/B}}{2}\right)^n}{n!} + \frac{(1-P_{\textrm{loss}}^{A/B}) \left(\frac{C_{e}^{A/B}}{2} \right)^{n-1}}{(n-1)!}, n \in \mathbb{N}^+.
    \end{cases}
\end{align*}
Consequently, we can write expression (\ref{eq:P_n1_n2}) for $P(n_1, n_2)$ as 
\begin{equation*}
    P(n_1, n_2) = e^{-\frac{C_{e}^{A} + C_{e}^{B} + 2 C_p}{2}} \left(I^{A}(0,n_1)  I^{B}(0,n_2) + C_p  I^{A}(1,n_1) I^{B}(1,n_2) \right).
\end{equation*}

It remains to find an expression for the probability of a 'good' coincidence click, i.e., a click originating from two source photons taking place in the same time-frame. This probability is given by the event that exactly one photon pair is produced, arrives in Alice's and Bob's labs, and is detected while no dark counts occur and no environmental photons meet Alice's and Bob's detectors, 
\begin{align*}
    P_{\textrm{Good}} = P_{\textrm{prod}}(1)(1-P_{\textrm{loss}}^A) (1-P_{\textrm{loss}}^B)\eta^A \eta^B (P_{\textrm{dark}}(0))^2\left(\sum_{k=0}^{\infty}P_{\textrm{env}}'^A(k)(1-\eta^A)^k\right)   \left(\sum_{k=0}^{\infty}P_{\textrm{env}}'^B(k)(1-\eta^A)^k\right).
\end{align*}

Finally, the isotropic noise parameter is given by $v = \frac{P_\textrm{Good}}{ P_{\textrm{TT}}(1,1)}$, such that
\begin{equation}
    \rho(v) = v \ketbra{\Psi_{\textrm{target}}} + (1-v) \frac{1}{d^2} \mathbbm{1}_{d^2 \times d^2}.
\end{equation}

In what follows, we assume that the photon source is placed in Alice's lab, which is in accordance with many practical realizations of QKD setups, such as the Free-Space Link between Vienna and Bisamberg. This means that Alice's source photons experience no channel loss and that there is no noise due to environmental photons on Alice's side, i.e. $P_{\textrm{loss}}^A = 0, C_e^A = 0$. 
The indicated realistic and practical setting leads to the following simplified expressions.

For Protocol 1 (where we place a polarization filter at the entrance of Bob's lab), we obtain
\begin{equation}
    \begin{aligned}
        P_{\textrm{TT}}(1,1) =& \frac{1}{2} e^{-2 C_{\textrm{d}}-C_{\textrm{p}}-\frac{C_{e}^{B}}{2}\eta_D} \\
        & \cdot\big[ C_{\textrm{p}}\eta_D^2\left( 2+C_{e}^{B}-2P_{\textrm{loss}}^B-C_{e}^{B}(1-P_{\textrm{loss}}^B)\eta_D \right) + C_{\textrm{d}}^2\left(2+2C_{\textrm{p}}(1-\eta_D)\left(1-\eta_D(1-P_{\textrm{loss}}^B) \right) \right)  \\
        &~~~~ +C_{\textrm{d}}\eta_D \left(C_{e}^{B}+C_{\textrm{p}} \left( 4-2P_{\textrm{loss}}^B-4\eta_D+4P_{\textrm{loss}}^B\eta_D+C_{e}^{B}(1-\eta_D \right)\left(1-(1-P_{\textrm{loss}}^B)\eta_D \right)  \right)  \big]
    \end{aligned}
\end{equation}
and 
\begin{equation}
    P_{\textrm{Good}} = C_{\textrm{p}}(1-P_{\textrm{loss}}^B) \eta_D^2 e^{-2C_{\textrm{d}}-C_{\textrm{p}}-\frac{C_{e}^{B}}{2} \eta_D},
\end{equation}
while for Protocol 2, we obtain

\begin{equation}
    \begin{aligned}
        P_{\textrm{TT}}(1,1) =& e^{-2 C_{\textrm{d}}-C_{\textrm{p}}-C_{e^{B}}\eta_D} \\
        & \cdot\big[C_{\textrm{d}}^2\left(1+C_{\textrm{p}}\right)+C_{\textrm{d}}\left( C_{e}^{B}+C_{\textrm{p}}\left( 2+C_{e}^{B}-C_{\textrm{d}}(2-P_{\textrm{loss}}^B)-P_{\textrm{loss}}^B\right) \right)\eta_D\\
        &~~+ C_{\textrm{p}} \left( 1+C_{e}^{B}-P_{\textrm{loss}}^B+C_{\textrm{d}}\left( C_{\textrm{d}}-2(1+C_{e}^{B})-C_{\textrm{d}}P_{\textrm{loss}}^B+(2+C_{\textrm{e}, B})P_{\textrm{loss}}^B \right) \right)\eta_D^2\\
        &~+(1-C_{\textrm{d}}) C_{\textrm{p}} C_{e}^{B}(-1+P_{\textrm{loss}}^B) \eta_D^3  \big]
    \end{aligned}
\end{equation}
and 
\begin{equation}
    P_{\textrm{Good}} = C_{\textrm{p}}(1-P_{\textrm{loss}}^B) \eta_D^2 e^{-2C_{\textrm{d}}-C_{\textrm{p}}-C_{e}^{B} \eta_D}.
\end{equation}

We note that this model covers all major contributions to white noise, which also represents the most dominant sources of noise and we leave more sophisticated noise models for future work.

\twocolumngrid

\end{document}